\input harvmac.tex
\input epsf
\noblackbox
%\draftmode
\def\figin{\epsfcheck\figin}\def\figins{\epsfcheck\figins}
\def\epsfcheck{\ifx\epsfbox\UnDeFiNeD
\message{(NO epsf.tex, FIGURES WILL BE IGNORED)}
\gdef\figin##1{\vskip2in}\gdef\figins##1{\hskip.5in}% blank space instead
\else\message{(FIGURES WILL BE INCLUDED)}%
\gdef\figin##1{##1}\gdef\figins##1{##1}\fi}
\def\DefWarn#1{}
\def\figinsert{\goodbreak\midinsert}
\def\ifig#1#2#3{\DefWarn#1\xdef#1{fig.~\the\figno}
\writedef{#1\leftbracket fig.\noexpand~\the\figno}%
\figinsert\figin{\centerline{#3}}\medskip\centerline{\vbox{\baselineskip12pt
\advance\hsize by -1truein\noindent\footnotefont{\bf Fig.~\the\figno } \it#2}}
\bigskipw\endinsert\global\advance\figno by1}
%
%
%
%
%\newcount\figno
% \figno=0
% \def\fig#1#2#3{
%\par\begingroup\parindent=0pt\leftskip=1cm\rightskip=1cm\parindent=0pt
% \baselineskip=11pt
% \global\advance\figno by 1
% \midinsert
% \epsfxsize=#3
% \centerline{\epsfbox{#2}}
% \vskip 12pt
% {\bf Fig.\ \the\figno: } #1\par
% \endinsert\endgroup\par
% }
% \def\figlabel#1{\xdef#1{\the\figno}}

\def\encadremath#1{\vbox{\hrule\hbox{\vrule\kern8pt\vbox{\kern8pt
 \hbox{$\displaystyle #1$}\kern8pt}
 \kern8pt\vrule}\hrule}}
 %
 %
 
 %%% Paragraphs

 %%% special math symbols
 \font\cmss=cmss10
 \font\cmsss=cmss10 at 7pt
 \def\rlx{\relax\leavevmode}
 \def\inbar{\vrule height1.5ex width.4pt depth0pt}
 \def\IC{\relax\,\hbox{$\inbar\kern-.3em{\rm C}$}}
 \def\IN{\relax{\rm I\kern-.18em N}}
 \def\IP{\relax{\rm I\kern-.18em P}}

\def\ZZ{\rlx\leavevmode\ifmmode\mathchoice{\hbox{\cmss Z\kern-.4em Z}}
  {\hbox{\cmss Z\kern-.4em Z}}{\lower.9pt\hbox{\cmsss Z\kern-.36em Z}}
  {\lower1.2pt\hbox{\cmsss Z\kern-.36em Z}}\else{\cmss Z\kern-.4em Z}\fi}
 %%% misc.
 \def\IZ{\relax\ifmmode\mathchoice
 {\hbox{\cmss Z\kern-.4em Z}}{\hbox{\cmss Z\kern-.4em Z}}
 {\lower.9pt\hbox{\cmsss Z\kern-.4em Z}}
 {\lower1.2pt\hbox{\cmsss Z\kern-.4em Z}}\else{\cmss Z\kern-.4em Z}\fi}
 %%% misc.
 \def\IZ{\relax\ifmmode\mathchoice
 {\hbox{\cmss Z\kern-.4em Z}}{\hbox{\cmss Z\kern-.4em Z}}
 {\lower.9pt\hbox{\cmsss Z\kern-.4em Z}}
 {\lower1.2pt\hbox{\cmsss Z\kern-.4em Z}}\else{\cmss Z\kern-.4em Z}\fi}

 \def\narrowplus{\kern -.04truein + \kern -.03truein}
 \def\narrowminus{- \kern -.04truein}
 \def\narrowminussub{\kern -.02truein - \kern -.01truein}

 \def\frac#1#2{{#1\over #2}}

 \def\IZ{\relax\ifmmode\mathchoice
 {\hbox{\cmss Z\kern-.4em Z}}{\hbox{\cmss Z\kern-.4em Z}}
 {\lower.9pt\hbox{\cmsss Z\kern-.4em Z}}
 {\lower1.2pt\hbox{\cmsss Z\kern-.4em Z}}\else{\cmss Z\kern-.4em Z}\fi}
 \def\IB{\relax{\rm I\kern-.18em B}}
 \def\IC{{\relax\hbox{$\inbar\kern-.3em{\rm C}$}}}
 \def\Ic{{\relax\hbox{$\inbar\kern-.22em{\rm c}$}}}
 \def\ID{\relax{\rm I\kern-.18em D}}
 \def\IE{\relax{\rm I\kern-.18em E}}
 \def\IF{\relax{\rm I\kern-.18em F}}
 \def\IG{\relax\hbox{$\inbar\kern-.3em{\rm G}$}}
 \def\IGa{\relax\hbox{${\rm I}\kern-.18em\Gamma$}}
 \def\IH{\relax{\rm I\kern-.18em H}}
 \def\II{\relax{\rm I\kern-.18em I}}
 \def\IK{\relax{\rm I\kern-.18em K}}
 \def\IP{\relax{\rm I\kern-.18em P}}
 %\def\IX{\relax{\rm X\kern-.01em X}}
 %this doesn't work

 \font\cmss=cmss10 \font\cmsss=cmss10 at 7pt
 \def\IR{\relax{\rm I\kern-.18em R}}

 %

 %
 %       \eqn\label{a+b=c}       gives displayed equation, numbered
 %                               consecutively within sections.
%     \eqnn and \eqna define labels in advance (of eqalign?)
 %
 \def\eqnn#1{\xdef
#1{(\secsym\the\meqno)}\writedef{#1\leftbracket#1}%
 \global\advance\meqno by1\wrlabeL#1}
 \def\eqna#1{\xdef
#1##1{\hbox{$(\secsym\the\meqno##1)$}}

\writedef{#1\numbersign1\leftbracket#1{\numbersign1}}%
 \global\advance\meqno by1\wrlabeL{#1$\{\}$}}
 \def\eqn#1#2{\xdef
#1{(\secsym\the\meqno)}\writedef{#1\leftbracket#1}%
 \global\advance\meqno by1$$#2\eqno#1\eqlabeL#1$$}

\newdimen\tableauside\tableauside=1.0ex
\newdimen\tableaurule\tableaurule=0.4pt
\newdimen\tableaustep
\def\phantomhrule#1{\hbox{\vbox to0pt{\hrule height\tableaurule width#1\vss}}}
\def\phantomvrule#1{\vbox{\hbox to0pt{\vrule width\tableaurule height#1\hss}}}
\def\sqr{\vbox{%
  \phantomhrule\tableaustep
  \hbox{\phantomvrule\tableaustep\kern\tableaustep\phantomvrule\tableaustep}%
  \hbox{\vbox{\phantomhrule\tableauside}\kern-\tableaurule}}}
\def\squares#1{\hbox{\count0=#1\noindent\loop\sqr
  \advance\count0 by-1 \ifnum\count0>0\repeat}}
\def\tableau#1{\vcenter{\offinterlineskip
  \tableaustep=\tableauside\advance\tableaustep by-\tableaurule
  \kern\normallineskip\hbox
    {\kern\normallineskip\vbox
      {\gettableau#1 0 }%
     \kern\normallineskip\kern\tableaurule}%
  \kern\normallineskip\kern\tableaurule}}
\def\gettableau#1 {\ifnum#1=0\let\next=\null\else
  \squares{#1}\let\next=\gettableau\fi\next}

\tableauside=1.0ex
\tableaurule=0.4pt

\def\IE{\relax{\rm I\kern-.18em E}}
\def\IP{\relax{\rm I\kern-.18em P}}

\Title
{\vbox{
 \baselineskip12pt
\hbox{CALT-68-2600, HUTP-06/A017}}}
 {\vbox{
 \centerline{On the Geometry of the String Landscape}
\centerline{} 
\centerline{and the Swampland}
 }}
\centerline{ Hirosi Ooguri$^1$  and  Cumrun Vafa$^2$}

\bigskip
\medskip 
\centerline{ $^1$California Institute of Technology}\smallskip
\centerline{Pasadena, CA 91125, USA}
\bigskip\centerline{ $^2$Jefferson Physical Laboratory, Harvard University}
\smallskip\centerline{Cambridge, MA 02138, USA}\smallskip

 \vskip .3in \centerline{\bf Abstract}

We make a number of conjectures about the geometry
of continuous moduli parameterizing the string landscape.
In particular we conjecture that such moduli
are always given by expectation value of scalar fields and that
moduli spaces with finite non-zero
diameter belong to the swampland.  We also conjecture
that points at infinity in a moduli space correspond
to points where an infinite tower of massless states appear, and
that near these regions the moduli space is negatively curved.
We also propose that there is no non-trivial 1-cycle of
minimum length in the moduli space. 
This leads in particular to the prediction 
of the existence of a radially massive partner to the axion. 
These conjectures put strong constraints on inflaton potentials
that can appear in a consistent quantum theory of gravity.
Our conjectures are 
supported by a number of highly non-trivial examples
from string theory. Moreover it is shown that these conditions can be violated if gravity is decoupled.  

 \smallskip \Date{May 2006}

\newsec{Introduction}
\lref\vswa{C.~Vafa,
  ``The string landscape and the swampland,''
  {\tt hep-th/0509212}.
  %%CITATION = HEP-TH 0509212;%%
}
\lref\douglasfinite{
  M.~R.~Douglas and Z.~Lu,
  ``Finiteness of volume of moduli spaces,''
  {\tt hep-th/0509224}.
  %%CITATION = HEP-TH 0509224;%%
}

\lref\armv{N.~Arkani-Hamed, L.~Motl, A.~Nicolis and C.~Vafa,
  ``The string landscape, black holes and gravity as the weakest force,''
  {\tt hep-th/0601001}.
  %%CITATION = HEP-TH 0601001;%%
}
\lref\kach{S.~Kachru, J.~McGreevy and P.~Svrcek,
  ``Bounds on masses of bulk fields in string compactifications,''
  {\tt hep-th/0601111}.
  %%CITATION = HEP-TH 0601111;%%
}
\lref\doug{M.~R.~Douglas, talk at the Strings 2005 Conference,
{\tt http://www.fields.utoronto.ca/} {\tt audio/05-06/strings/douglas/}.}
The fact that string theory seems to offer a diverse range of possibilities
for vacua has been viewed as a drawback for the theory:  We cannot converge
on a precise prediction for the theory.  However despite this
diversity of options for the string landscape, it has been pointed out in
\vswa\ that there are also a number of patterns that seem to emerge.  Not
every effective field theory that appears consistent seems to arise in string
theory.   It is natural to conjecture that these theories are not fully
consistent as quantum gravitational theories.
 Such theories belong to the swampland.   
The condition
to be on the swampland can be lifted if we decouple gravity.  In other words
 a fully consistent quantum
field theory cannot always be coupled to gravity.
 It is
natural to conjecture that points on the swampland are ``anomalous quantum
gravitational theories,'' which are anomalous
in a more subtle way, than has been discovered in the context of quantum
field theories.
It would be important to improve the restrictions for theories to
 arise in string theory.  By tabulating such restrictions
and finding the criteria that distinguish the swampland from the landscape
one can hope to have a deeper understanding into the universality
class of quantum gravitational theories.  The main aim of this paper
is to take some modest steps in this direction.  For some related conjectures
distinguishing swampland from the landscape see \refs{\douglasfinite,\armv,
\doug, \kach}.

One conjectural criterion to be on the string landscape is that
the volume of the
moduli space seems typically finite. 
There are, however, well known counter-examples
to this seemingly general phenomenon. Consider compactification on a 
circle of radius $r$. The moduli space for the circle has a metric 
\eqn\metric{ ds^2 =\left( {dr\over r}\right)^2,}
and the volume integral is divergent. In \vswa,  
it was pointed out that this volume divergence correlates with the 
cutoff in mass. Namely, let $\epsilon$ be a fixed scale for the low energy
effective theory and that we insist that all the higher massive scales 
have mass greater than $\epsilon$. For large $r$ and small $\epsilon$, the region of 
the moduli of the moduli space satisfying this constraint is
specified by $r < 1/\epsilon$. Clearly the 
volume of this region of the moduli space is finite, 
$$\int^{1/\epsilon} {dr \over r} = - \log \epsilon .$$
(The lower bound for the $r$ integral should also be regularized
in a similar fashion, as will be clear in our examples.) We find the logarithmic volume divergence as we take
the limit $\epsilon \rightarrow 0$. Thus, in this case, the volume
divergence is related to the emergence of infinitely many 
extra light particles.

Even though this divergence might seem special to 1-dimensional
moduli spaces, here we wish to formulate conjectures applicable
to {\it every} moduli space ${\cal M}$ encountered in string theory.
Our conjectures apply both to the moduli
space of scalars, as well as the subspaces parameterizing minimum loci
for the potentials defined on such spaces.  In this
way our conjectures can be viewed as very powerful constraints on what
potentials can appear in a consistent theory of quantum gravity.

Our conjectures suggest the following picture for the moduli
space of a consistent quantum gravitational theory:  The moduli
space is parameterized by expectation value of
scalar fields.  We conjecture that there are points
infinitely far away from one another on the moduli space and that
the points near infinity are points where a tower of massless modes
appear (with exponentially small mass as a function of the distance to such points). 
Infinite distance singularities combined with finite volume typically imply 
negative curvatures. Thus, we conjecture that the curvature becomes negative 
near points at infinity.   
We also propose that there are no non-trivial loops of
minimum length in the moduli space of scalars.
As we will elaborate later this is in line with the 
intuition that the duality groups are generated by discrete gauge symmetries
realized at different points on the moduli space.

The organization of the paper is as follows:  In section 2 we present the precise
form for our conjectures.  In section 3 we present examples of how
string theory supports our conjectures and how if we decouple gravity we
can violate them.  In section 4 we discuss some field theoretic
considerations related to our conjectures.  In section 5 we end with 
some concluding thoughts.

\newsec{The Conjectures}

In this section we present our conjectures.
We claim that our conjectures apply to consistent quantum
theories of gravity with finite Planck mass in
4 and higher spacetime dimensions. We do not consider 3
or lower dimensions as gravity does not contain propagating
degrees of freedom in these dimensions, though some of our
conjectures may be applicable to 3 dimensional cases as well. 
Our conjectures should apply to string theories compactified to 4 and
higher dimensions. But they should also apply to other consistent quantum
gravity theories in these dimensions if they exist. 

Let ${\cal M}$ denote the moduli space of a consistent quantum
gravity.  Choosing a point $p\in {\cal M}$ corresponds
to fixing the low energy effective Lagrangian for the theory.
Different points in ${\cal M}$ are in different super-selection
sectors. However, we could ask if we can vary $p$ {\it locally} in spacetime
while keeping the asymptotic boundary conditions fixed at $p$.
Our conjecture states that this is possible.
In addition, if we consider a finite volume compactification down
to 0 or 1 spatial dimensions 
(say on a torus), then we can vary the point $p$ and study 
${\cal M}$. Our first conjecture
states that ${\cal M}$ is given by the expectation value of scalar
fields.  

\medskip
\noindent
{\bf Conjecture 0:}  ${\cal M}$ is parameterized by inequivalent expectation
values of massless scalar fields.

\medskip
\noindent
This conjecture in the context of string theory is so well known that
one does not even bother to state it:  {\it There are no coupling
constants in string theory}; every parameter can be varied by changing
the expectation value of a field.  This did not have to be the case and
one could a priori imagine a consistent quantum gravitational theory for
which the parameters are fixed (and in particular cannot be varied locally).
We conjecture that such gravitational theories cannot exist as a
fully consistent quantum gravitational theory and belong
to the swampland. 

\medskip
Given that points on ${\cal M}$ are given by the expectation value
of scalar fields, we can use their kinetic
term to define a metric on ${\cal M}$.  We can then
state our next conjecture dealing with the geometry
of ${\cal M}$:  Let us assume that the dimension of ${\cal M}$ is
not zero.
Let us define $d(p_1, p_2)$ for two points $p_1, p_2 \in {\cal M}$
as the distance of the shortest geodesic between them measured in Planck units. 
We conjecture that the distance can take an arbitrary large
value, namely:

\medskip
\noindent
{\bf Conjecture 1:} Choose any point $p_0 \in 
{\cal M}$. For any positive $T$, there is another point 
$p \in {\cal M}$ such that $d(p, p_0) > T$.  

\medskip
\noindent
More strongly, we claim that the limit $d(p, p_0) \rightarrow \infty$
is correlated with a breakdown of the low energy effective theory
in a particular way:
 
\medskip
\noindent
{\bf Conjecture 2:} Compared to the theory at $p_0 \in {\cal M}$, 
the theory at $p$ with $d(p, p_0) > T$ has an infinite tower of light
particles starting with mass of the order of  $e^{-\alpha T}$
for some $\alpha > 0$.  In the $T\rightarrow \infty$ limit, the number of extra light 
particles of mass less than a fixed mass scale 
becomes infinite.  

\medskip
\noindent
In the first part of the conjecture 2, we are claiming that
the low energy effective theory defined on a particular point
on the moduli space makes sense only in a domain in a finite
diameter from the point. In section 4 we will give a physical
motivation for this part of the conjecture. If the breakdown
of the low energy theory is associated with an infinite number
of extra light particles, one-loop graphs of such particles can give
rise to the distance divergence. If only a finite number of 
particles becomes light, they alone do not generate the distance
divergence for dimensions higher than 2. This motivates the
second part of the conjecture 2.  Note that this conjecture in particular places strong constraints
for inflationary models:  We cannot have a slow roll inflation where the distance
in the scalar moduli space is much bigger than Planck length and still use
the same effective field theory.

\medskip
\noindent
{\bf Conjecture 3:} The scalar curvature near the points
at infinity is non-positive. (It is strictly negative if
the dimension of the moduli space is greater than 1.)  

\medskip
\noindent
The fact that the curvature
is negative near infinity is correlated with the fact that there are points
infinitely far away, even though the volume is finite.  

One could consider a stronger conjecture claiming 
various sectional curvatures being negative near infinity.
The geodesics deviation equation shows that neighboring geodesics
tend to diverge when curvatures are negative. In particular,
in a compact space, this leads to mixing of geodesics.  
More precisely, it is known that the geodesics flow 
in a compact space with negative sectional curvatures
is ergodic \ref\anosov{D.~V.~Anosov,
``Geodesic flows on closed Romanian manifolds with
negative curvature,'' Proc. Steklov Inst. Mathematics {\bf 90}, 209 (1967). }. 
Our conjectures here go in that direction.
However they are not quite the same condition, as for instance, there are
points on ${\cal M}$ arising in string theory where the sectional curvature
is positive, as will be noted in the next section.  Our conjecture here only
refers to the scalar curvature being negative near infinity, which is weaker.
Nevertheless, it could be true that the geodesic motion on the moduli space
is ergodic.  This could be potentially true, independently of the above conjecture,
and would be worth investigating.

\medskip
\noindent
{\bf Conjecture 4:} There is no non-trivial 1-cycle with
minimum length within a given homotopy class in ${\cal M}$. 

\medskip
\noindent
%The motivation of this conjecture is twofold.  
Suppose
${\cal M}$ is obtained from dividing a contractible Teichm\"uller ${\cal T}$
space by
a group action $\Gamma$, as is often the case in string theory:
$${\cal M}={\cal T}/ \Gamma .$$
Then the duality group associated to ${\cal M}$ is $\Gamma$.  Fixed points
of $\Gamma$ have a nice physical significance:
If $g\in \Gamma$
fixes a point $p\in {\cal T}$, then $g$ is a gauge symmetry at $p$.  It seems
to be the case that $\Gamma$'s that appear in string theory are generated by
group elements which have fixed points.  In other words the entire $\Gamma$ can
be viewed as a gauge symmetry which is broken as we move around in ${\cal M}$, and
we recover different parts of it at different points of ${\cal M}$.  If $\Gamma$
is indeed generated by elements with fixed points, then $\pi_1({\cal M})$ would be
trivial, as each loop which can be identified with an element $h\in \Gamma$
can be written as $h=\prod g_i$ and each $g_i$ has fixed points and so
the corresponding segment of the path can be contracted to a point.  So the entire
path would be contractible, thus making ${\cal M}$ simply connected.
Clearly in this case there is no 1-cycle of minimum length.

Another motivation for this conjecture is the following. 
Suppose we compactify the theory further down to (1+1) dimensions
and choose the space to be a circle. If there is a non-trivial 
1-cycle in ${\cal M}$ with minimum length, we can use it to define 
a state with a non-zero global charge by going over the corresponding 
path on the circle.  Since there are no global charges in a gravitational 
theory (to be consistent with no-hair theorems for black holes),
this cannot happen.  This suggests that at least the center of $\pi_1$ is trivial.

Conjecture 4 suggests there is a partner to the axion. 
The axion is massless before we take into account
instanton effects, and its has a compact moduli space of $S^1$. 
If we can separate the scale of the instanton effects from
those at the Planck scale so that the instanton effects are
taken into account in the low energy effective theory, the 
conjecture 4 would require that there exists 
another direction where $S^1$ shrinks to a point,
$i.e.$ a radial partner to the axion. 

\newsec{Evidences for the Conjectures}

In the following, we will present evidences for these conjectures.  
We will also present examples in string theory that, if we
decouple gravity by taking the internal volume to be infinitely
large, then these constraints disappears.  So it is only a constraint
in the context of quantum theories of gravity. 
%Below 
%we first consider examples where gravity is dynamical.  We in particular
%consider compactifications of M-theory on $S^1$, type IIB
%in 10 dimensions, compactifications of
%IIA,B on $S^1$, compactifications of heterotic/type I string
%on $S^1$, compactifications of type II strings on $K3$ and also on Calabi-Yau
%threefolds.  We then consider examples of how this property fails when
%gravity is turned off.  Examples will include field theories living on
%D-branes in $R^{10}$, where no massless
%modes are encountered on complete geodesics,
%and type II strings propagating
%on certain non-compact Calabi-Yau manifolds with wrapped branes, where
%$\pi_1({\cal M})\not =0$.  In this sense our conjectures are sharp.

\subsec{Gravity Examples}

\smallskip

\noindent
{\it Example}  i) \ M-theory Compactified on $S^1$

\medskip

In the 10-dimensional theory, which is equivalent to type IIA string,
we have a massless scalar $r$ whose vev gives the radius of the circle. 
Since the metric on this space is $(dr/r)^2$, 
the geodesic distance
between $r$ and $r_0$ is given by
$$ T = | \log(r/r_0)|. $$
Since $r$ and $r_0$ can take arbitrary positive values, 
the conjecture 1 is clearly satisfied. 

If we fix $r_0$ and take $T \rightarrow \infty$, 
we have either $r \rightarrow \infty$ or $0$.  
On the one hand, in the limit of $r \rightarrow \infty$, 
the M theory develops light Kaluza-Klein modes 
with mass $1/r$ in the 11-dimensional frame.
In the Einstein frame in 10 dimensions, which involves rescaling
the metric by $r^{1/4}$, we find the lowest mass scale $\epsilon(T)$ is 
$$\epsilon( T ) \sim r^{-1/8} r^{-1} \sim {\rm exp}
\left(-{9\over 8}T\right)$$
On the other hand, for the compactification
at $r = r_0 e^{-T}$, the radius $r\rightarrow 0$ and
the theory again develop light states, namely type IIA strings, 
corresponding to membranes wrapping $S^1$. The tension of 
the string will scale as  $r$, which means that their mass 
in 11-dimensional frame scales as $r^{1/2}$. Or in 10-dimensional 
Einstein frame as
$$\epsilon(T)\sim r^{-1/8} r^{1/2} =r^{3/8} \sim
{\rm exp}\left(-{3\over 8}T\right)$$
So we see that this result is consistent with the conjecture 2
since in both cases  
infinitely many light modes appear as $T \rightarrow \infty$.  Note that as this example
demonstrates
the coefficient of the exponent is not universal 
(9/8 and 3/8 in this case) even for a given dimension.

\bigskip

\noindent
{\it Example} ii) \ Type IIB Strings in 10 Dimensions

\medskip

The dilaton-axion moduli space is $SL(2,{\bf Z})\backslash 
SL(2,{\bf R}) / SO(2)$, parametrized by $\tau = \tau_1 + i\tau_2$,
where $\tau_1$ and $\tau_2$ are related to vev's of the axion
and the dilaton. The metric in this space is
\eqn\poincaremetric{ ds^2 = {d\tau d\bar\tau \over ({\rm Im}\tau)^2}. }
This is a good example to illustrate the point
of our conjectures since the volume of this moduli space is finite
yet a geodesic length toward $i \infty$ is logarithmically
divergent. As we take $\tau \rightarrow i\infty$ keeping
$\tau_0$ finite, the geodesic length between $\tau$ and
$\tau_0$ is approximately given by
$$ T \sim \log({\rm Im}\ \tau/{\rm Im}\ \tau_0). $$ 
Since light stringy excitations of
mass $\sim e^{-{1\over 4}T}$ appear at $\tau$, 
this exemplifies the conjectures 1 and 2.  

The metric \poincaremetric\ has a constant negative curvature, consistently
with the conjecture 3. The curvature integral as well as the volume 
integral is finite,
thanks to the quotienting by the S-duality symmetry $SL(2,{\bf Z})$.
It is known that the geodesic flow on compact spaces with 
constant negative curvatures is ergodic \ref\cha{
E. Hopf, Leipzig Ber. Verhandl. S\"achs. Akad. Wiss.
{\bf 91}, 261 (1939); 
For discussion of 
the ergodicity of geodesic flows 
in the context
of string compactification, see G.~W.~Moore, 
  ``Finite in all directions,'' 
  {\tt hep-th/9305139}. 
  %%CITATION = HEP-TH 9305139
}. 
In particular, if we consider a region 
$\tau_2 > 1/\epsilon$ in the fundamental domain,
generic geodesics will pass through it within 
finite geodesic time.

This moduli space is simply connected since $SL(2,{\bf Z})$
is generated by $S$ and $ST$ transformations ($S: \tau 
\rightarrow -1/\tau, ~T: \tau \rightarrow \tau +1$), both of which
have fixed points in the fundamental domain,
at $\tau=i$ and $\tau ={\rm exp}(2\pi i/3)$
respectively. 

\bigskip

\noindent
{\it Example} iii) \ Type IIB Strings Compactified on $S^1$  

\medskip

In this case the moduli space of compactification is characterized
by the radius $r$ of $S^1$, in addition to the $\tau$ space discussed
above.  This example is similar to the example i, and we have
$$T = |\log (r/r_0)| $$
in string frame in 10 dimensions.  We get light 
Kaluza-Klein modes for $r \rightarrow \infty$ and light winding modes for
$r \rightarrow 0$.  In the 9-dimensional Einstein frame, 
this translates to the existence of light mass 
scales for large $T$ given by
$$\epsilon(T)\sim r^{-1/7} \exp(-T)\sim 
\exp\left( -{8 \over 7} T\right)
~~{\rm or}~~\exp\left( - {6 \over 7} T \right)$$
Thus, both conjectures 1 and 2 are satisfied. 
%\VafaIH
\lref\VafaIH{
  C.~Vafa,
 ``Quantum symmetries of string vacua,''
  Mod.\ Phys.\ Lett.\ A {\bf 4}, 1615 (1989).
  %%CITATION = MPLAE,A4,1615;%%
}
\bigskip

\noindent
{\it Example} iv) \  Theories with 16 Supercharges

\medskip

The classical moduli space of $K3$ is
$$ {\cal M}_{K3}^{(classical)} = SO(3,19; {\bf Z})\backslash
SO(3,19)/(SO(2) \times SO(19)) . $$
This is the relevant moduli space for the 7 dimensional theory
obtained by compactifying M-theory on $K3$.
The physical moduli
space of type II string on $K3$ includes configurations of
the NS-NS 2-form, the dilaton, and the R-R forms.
In type IIA theory, the resulting moduli space, in addition
to the coupling constant (as discussed 
in the example iii in the above), is
$$ {\cal K}_{K3}^{IIA} = SO(4,20; {\bf Z})\backslash
SO(4,20)/(SO(4) \times SO(20)). $$
In type IIB theory, the coupling constant multiplet
and other fields combine to give
$$ {\cal M}_{K_3}^{IIB} = SO(5,21; {\bf Z})\backslash
SO(5,21)/(SO(5) \times SO(21)). $$
In each of these moduli spaces, any point which
is infinite distance away from points in the middle
of the moduli space is dual to a decompactification
limit. Therefore an infinite tower of light Kaluza-Klein
modes appears at each of these points, in accord with
the conjectures 1 and 2.   

These spaces have negative curvatures, satisfying
the conjecture 3. In general, 
if $\Gamma$ is a lattice in a connected 
semi-simple Lie group $G$ with finite center
and 
$K$ is a  maximal compact subgroup of $G$, 
generic geodesics are dense 
in the coset space $\Gamma \backslash G / K$
\ref\moore{C.~C.~Moore, 
``Ergodicity of flows on homogeneous spaces,''
Amer. J. Math., 88 (1966) 154.}\ leading to ergodicity. 

\lref\AspinwallRG{
  P.~S.~Aspinwall and D.~R.~Morrison,
  ``String theory on K3 surfaces,''
  {\tt hep-th/9404151}. 
 %%CITATION = HEP-TH 9404151;%%
}
  
For the $K3$ moduli space, the discrete group $\Gamma$
is known to be generated by symmetry of the worldsheet
CFT's \refs{\VafaIH, \AspinwallRG}, and this supports the expectation from the conjecture 4
that these moduli spaces are simply connected. 

The same is true for heterotic/type I string compactified
on $T^d$, whose moduli space (in addition to the coupling constant) is
$$  SO(d,16+d; {\bf Z})\backslash SO(d,16+d; {\bf R})/ (SO(d)
\times SO(16+d)). $$

\bigskip
\noindent
{\it Example} v) \ Compactifications of Type II Strings on Calabi-Yau
threefolds   

\medskip

The moduli space splits into the vector multiplet moduli space, 
which has special K\"ahler structure, and the hypermultiplet moduli
space, which is quaternionic K\"ahler. In the type IIB theory,
the vector multiplet moduli space is parametrized complex 
structure of the Calabi-Yau manifold;  in the type IIA theory,
it is parametrized by complexified K\"ahler structure.
The K\"ahler moduli space has at least one infinite distance
singularity corresponding to the decompactification limit,
where infinitely many Kaluza-Klein modes become light. By
the mirror symmetry, we expect that the complex moduli space has an infinite distance 
singularity. This demonstrates the conjecture 1. 

The curvature of the vector multiplet moduli space
is not necessarily negative. However, one can see from explicit
examples as that in \ref\candelas{ P.~Candelas, 
X.~C.~De La Ossa, P.~S.~Green and L.~Parkes,
  ``A pair of Calabi-Yau manifolds as an exactly soluble superconformal
  theory,''
  Nucl.\ Phys.\ B {\bf 359}, 21 (1991).
  %%CITATION = NUPHA,B359,21;%%
} that the scalar curvature asymptotes to negative
toward infinite distance singularities. In fact,
this is a general fact. By using results in
\ref\Lu{
  Z.~Lu and X.~Sun,
  ``On the Weil-Petersson volume and the first Chern class of the moduli space
  of Calabi-Yau manifolds,''
  Commun.\ Math.\ Phys.\  {\bf 261}, 297 (2006);
  {\tt math.dg/0510021}.
  %%CITATION = MATH-DG 0510021;%%
}, one can estimate the behavior
of the metric near singularities. Since the moduli space is orientable, 
the Einstein action can be written
as
$$ \int_{{\cal M}} \sqrt{g} R \sim \int c_1 \wedge k^{n-1}, $$
where $c_1$ is the first Chern class, $k$ is the K\"ahler form,
and $n$ is the complex dimension of ${\cal M}$. This can
be expressed as an integral of the same quantity of 
a compactification $\overline{\cal M}$ of ${\cal M}$, which
is topological, plus contributions from singularities. 
One can show that infinite distance singularities always
give strictly negative contributions to the integral,\foot{We would 
like to thank Z. Lu for helping us make this estimate.} as expected from
the conjecture 2.   For Calabi-Yau manifolds with 
one K\"ahler moduli, the large
size limit is always given by constant negative curvature (as the prepotential
is dominated by $X_1^3/X_0$ in the standard notation).

Are Calabi-Yau moduli spaces simply connected?  
One way to test this is to see whether 
monodromies of period integrals are
generated by transformations with fixed points 
represented by Calabi-Yau manifolds with symmetries. 
One subtlety is that such points can be at infinite
distance away from the middle of the moduli space,
in which case one has to be careful about how to
define $\pi_1$. 
Indeed there is an examples of where the moduli space is 
complex one-dimensional with two different large 
complex structure limits \ref\fundamental{E.~A.~Rodland, 
``The Pfaffian Calabi-Yau, its mirror, and their link to the Grassmannian $G(2,7)$,''
Compositio Math. {\bf 122}, 135 (2000).}.\foot{We thank 
D. Morrison for bringing this example into our attention.}
 If we remove 
the two points in the limits, the moduli space has the topology of
${\bf C}^*$ and is not simply connected. 
In this case, the length of a circle that goes around
the non-trivial $\pi_1$ of ${\bf C}^*$ vanishes as the
circle approaches to either one of the large complex structure 
limits. This follows from the fact that the limits are
infinite distances away yet the volume of the moduli space
is finite. Therefore the conjecture 4 is still valid in this
case.

As for the hypermulitplet moduli space, though a quaternionic
K\"ahler manifold in general can have curvatures of any sign \ref\wolf{J. A. Wolf, ``Complex 
homogeneous contact manifolds
and quarternionic symmetric spaces,'' J. Math. Mech. {\bf 14}, 1033 (1965).},
it is known that those realized for hypermultiplet fields coupled
to supergravity must have a negative scalar curvatures
\ref\baggerwitten{
 J.~Bagger and E.~Witten,
  ``Matter couplings in ${\cal N}=2$ supergravity,''
  Nucl.\ Phys.\ B {\bf 222}, 1 (1983).
  %%CITATION = NUPHA,B222,1;%%
}, satisfying the conjecture 2. Note that the negativity of the
scalar curvature in this case is a direct consequence of the
local ${\cal N}=2$ supersymmetry, and therefore it applies to 
the hypermultiplet moduli space after perturbative and non-perturbative
quantum corrections are taken into account as far as we consider
the low energy effective theory with minimum number of derivatives. 

\bigskip
\noindent
{\it Example} v) \ Compactifications to Four Dimensions with
${\cal N}=1$ Supersymmetry

\medskip
Little is known about the metric on ${\cal M}$ in this case. 
To our knowledge, the conjectures 1, 2, and 4 are consistent with
all known examples. (Without information on the metric, 
it is difficult to test the conjecture 3 about the curvature
of ${\cal M}$.) Consider for example the flux compactifications in 
\ref\DeWolfeUU{
  O.~DeWolfe, A.~Giryavets, S.~Kachru and W.~Taylor,
  ``Type IIA moduli stabilization,''
  JHEP {\bf 0507}, 066 (2005);
  {\tt hep-th/0505160}.
  %%CITATION = HEP-TH 0505160;%%
}. In this construction, the classical superpotential 
generated by RR and NS-NS fluxes fixes all of the K\"ahler 
moduli and the complex structure moduli. There is one 
axion for each complex structure moduli multiplet that 
cannot be stabilized by the fluxes.
In some models, these axions remain massless to 
all orders in $\alpha'$ and $g_s$ expansions.
However, in all known examples of this type, 
there are always instantons of the appropriate
charges to lift all the remaining axions.\foot{We thank
S. Kachru for discussion on this point.} 
Thus, it appears that there are obstructions in 
constructing models with compact moduli spaces
with finite diameter ($i.e.$ with a upper bound
on $d(p,p_0)$) or with non-trivial fundamental group,
and we regard this difficulty as in support of
the conjectures 1, 2, and 4.

\subsec{Non-Gravitational Counter-Examples}

We claim that the conjectures hold for theories with gravity
with finite Planck scale. 
It is not difficult to construct 
non-gravitational field theory models which violate
the conjectures.  In this sense our conjectures focus on what aspects
of a consistent quantum theory are required purely from requiring
having gravity.
 For example, if the theory has a compact global symmetry
$G$, as is typically the case in quantum field theories,
and if it is spontaneously broken to a subgroup $H$, the moduli 
space $G/H$ has finite radius and the conjecture 1 fails in this
case. 
Such spaces can have non-trivial fundamental groups, violating
the conjecture 4 also.  
Since string theory cannot have any continuous global symmetry, 
such an example can exist only in the limit where we decouple
gravity. In the following, we will present examples
to show how the violation of the conjectures is explicitly 
tied to the decoupling of gravity. 

\bigskip
\noindent
{\it Examples} a) \ Field Theories Living on Flat D-Branes in $R^{10}$.

\medskip

The gauge theory on flat parallel D$p$ branes in $R^{10}$ contains
$(9-p)$ massless scalar fields in the adjoint representation of 
the gauge group. If we go to a generic direction away from the 
origin of the moduli space, we can go to infinity along geodesic 
without encountering extra massless particles. Therefore,
this gives an example where the distance $d(p,p_0)$ can
become infinite (and therefore the conjecture 1 holds), 
but no new massless particles appear there (and the conjecture 2 fails).
In this case, the gravity degrees of freedom in the bulk 10 dimensions
is decoupled. To have a finite Planck scale in $(p+1)$ dimensions
along the branes, we need to compactify the transverse direction
appropriately. This will change the structure of the gauge theory
moduli space. 

\bigskip
\noindent
{\it Example} b) Rigid Limit of ${\cal N}=2$ Theories in Four Dimensions

\medskip

We saw in the example v in section 3.1 that, in theories
with local ${\cal N}=2$ supsersymmetries, the
sign of curvature of vector multiplet moduli may be indefinite, 
but the scalar curvature becomes strictly negative near
infinite distance singularities. 

On the other hand, it is straightforward to see that, 
in non-gravitational theories with global ${\cal N}=2$
supersymmetry, the Ricci curvature of the vector multiplet moduli
space is positive definite. In this case, the metric can be written
as
$$ g_{i\bar j} = {\rm Im}\ \tau_{ij}, ~~~~\tau_{ij} =
\partial_i \partial_j F(z), $$
where $F(z)$ is the pre-potential. The Ricci curvature then is
$$\eqalign{ R_{i\bar j} &= - \partial_i \bar\partial_{\bar j} \log \det g
\cr & = {1 \over 4} g^{k\bar l} g^{m\bar n} 
(\partial_i \tau_{km})(\bar \partial_{\bar j}\bar\tau_{\bar l \bar n}),}$$
which is manifestly positive definite. The essential step is
to use the fact that $g_{i\bar j}$ is the imaginary part of $\tau_{ij}$,
which is holomorphic in $z$. 

It is instructive to see how the decoupling of the gravity leads
to the positive definite Ricci curvature. In a theory with
local ${\cal N}=2$ supersymmetry, the K\"ahler potential $K$
for the moduli space metric is given by %\deWitPK
\ref\deWitPK{
  B.~de Wit and A.~Van Proeyen,
  ``Potentials and symmetries of general gauged 
${\cal N}=2$ supergravity - Yang-Mills models,''
  Nucl.\ Phys.\ B {\bf 245}, 89 (1984).
  %%CITATION = NUPHA,B245,89;%%
}
$$ K = - \log \left( 4 {\cal F} - 4 \bar {\cal F} 
+ \bar{z}^{\bar i} \partial_i {\cal F}
                     - z^i \bar\partial_{\bar i} \bar{\cal F} \right), $$
where ${\cal F}(z)$ is the pre-potential of the gravitational theory.
The rigid limit can be taken by setting
$$ {\cal F}(z) = {i\over 8} M_{Pl}^2 + F(z), $$
and let $M_{Pl} \rightarrow \infty$. The K\"ahler potential then becomes,
modulo terms that are purely holomorphic or anti-holomorphic,
$$ K = - \log iM_{Pl}^2 + 
{1\over i M_{Pl}^2} \left( \bar{z}^{\bar i} \partial_i F(z)
- z^i \bar{\partial}_{\bar i} \bar{F}({\bar z})\right) 
 + \cdots. $$
in this limit, the metric $g_{i\bar j}$ becomes
an imaginary part of the holomorphic $\tau_{ij}$ and the Ricci
curvature is positive definite. This shows how the conjecture 2
can be violated by taking the decoupling
limit $M_{Pl} \rightarrow \infty$.

\bigskip
\noindent
{\it Example} c) \  Type II Strings on Non-Compact Calabi-Yau Manifolds 
with wrapped branes

\medskip

Another way to obtain field theories decoupled from gravity
is to consider D-branes on cycles in a non-compact Calabi-Yau
manifold $M$ and to take the low energy limit. For example,
consider a brane at a point in $M$. Though the
moduli space metric is not exactly equal to that of $M$
\lref\dos{
M.~R.~Douglas, H.~Ooguri and S.~H.~Shenker,
  ``Issues in (M)atrix model compactification,''
  Phys.\ Lett.\ B {\bf 402}, 36 (1997);
  {\tt hep-th/9702203}.
  %%CITATION = HEP-TH 9702203;%%
} 
\lref\do{
M.~R.~Douglas and H.~Ooguri,
  ``Why matrix theory is hard,''
  Phys.\ Lett.\ B {\bf 425}, 71 (1998);
  {\tt hep-th/9710178}.
  %%CITATION = HEP-TH 9710178;%%
} \refs{\dos,\do}, 
the metric computed using the gauge theory method
of 
%\DouglasSW
\ref\DouglasSW{
  M.~R.~Douglas and G.~W.~Moore,
  ``D-branes, quivers, and ALE instantons,''
  {\tt hep-th/9603167}.
  %%CITATION = HEP-TH 9603167;%%
} is typically Ricci flat 
\ref\DouglasZJ{
  M.~R.~Douglas and B.~R.~Greene,
  ``Metrics on D-brane orbifolds,''
  Adv.\ Theor.\ Math.\ Phys.\  {\bf 1}, 184 (1998);
  {\tt hep-th/9707214}.
  %%CITATION = HEP-TH 9707214;%%
}. In particular, its scalar curvature does not 
necessarily become negative as we go infinite distance
away from the origin of $M$.

As another example, consider a non-compact local model of Calabi-Yau
manifold with small ${\bf P}^2$. The moduli space of a
brane wrapping  a 2-cycle in ${\bf P}^2$ is  
a copy of ${\bf P}^2$ (see \ref\kkv{ S.~Katz, A.~Klemm and C.~Vafa,
``M-theory, topological strings and spinning black holes,''
  Adv.\ Theor.\ Math.\ Phys.\  {\bf 3}, 1445 (1999);
{\tt hep-th/9910181}.}\ for a study
of 2 branes in this geometry).
Note that this space is compact (violating conjecture
1) and positively curved (violating conjecture 3).
Note that cycles can shrink in a Calabi-Yau manifold only when
they are positively curved. This may be related to the fact
that scalar moduli coupled to gravity tend to be negatively
curved, as suggested in the conjecture 3. 

 \bigskip
\noindent
{\it Example} d) \ D-brane Wrapping $T^2$

\medskip

Consider type IIB compactified on $T^2$ to 8 dimensions.  Consider
a D5 brane wrapping $T^2$ and filling a 3+1 dimensional flat
subspace of 8 dimensions.  The theory living on this 4 dimensions is
a deformed ${\cal N}=4$ Yang-Mills, where out of the 6 scalars, two of
them (corresponding to the Wilson lines on $T^2$) are periodic.  The
theory on this brane does not have a dynamical gravity in 4 dimensions
(as the transverse dimensions are non-compact) and we have the moduli
space of the theory given by
$$R^4\times T^2$$
This clearly violates conjecture $4$ which states that the moduli space
should have no non-trivial loops.  $T^2$ has non-trivial loops.

\newsec{Motivation from Effective Field Theories}

We would like to point out that there are phenomena in 
a low energy effective theory coupled to gravity that 
are closely related the property of moduli spaces discussed in the
above. 

\subsec{Infinite Distances and Appearance of a Tower of Light States}

The conjecture 2 implies that there is an infinite tower of light particles 
at infinite distance from any point inside the moduli space,
where the effective field theory in the interior breaks down, and potentially
a new description takes over. This is consistent
with the following field theory fact. 

Suppose there is a field $\varphi$, whose mass $m(\phi)$ is a function
of the moduli field $\phi$. We assume that $m(\phi)$ is generically large
in the moduli space ${\cal M}$ so that we can integrate out $\varphi$. Now suppose
there is a particular point $\phi_0 \in {\cal M}$ where $m(\phi_0)=0$. 
There the effective field theory with $\varphi$ integrated out breaks down. 
To see its effect on the moduli space metric, let us look at how the one-loop
diagram of the $\varphi$ field modifies the kinetic term of $\phi$. 
If $\varphi$ is a scalar field, the coupling between $\varphi$ and
$\phi$ is given by ${\cal L}_{int} = m(\phi)^2 \varphi^2$, and it gives
the one-loop correction:
$$ g_{ij}^{{\rm one-loop}} = 
  m(\phi)^{d-4} {\partial m \over \partial \phi^i} {\partial m \over
  \partial \phi^j} .$$
Here $d$ is the spacetime dimensions of the Minkowski space where
we are considering the gravity theory.  Consider a point $\phi_1$
near $\phi_0$ where $m(\phi_1)= \epsilon \ll 1$ (in Planck units).
The distance to $\phi_1$
measured by the quantum corrected metric in the above is given by
$$ d(\phi, \phi_1) = \int_{\epsilon} m^{{d-4\over 2}} dm
         = - {2\over d-2} \epsilon^{{d-2\over 2}} + {\rm finite}. $$
Thus, a finite number of light particles does not generate
infinite distance for $d > 2$.  
We get the same result if the light particle is a fermion
$\psi$ with an interaction ${\cal L}_{int} = m(\phi) \bar\psi \psi$.  

On the other hand, a sum over one-loop effects of an infinite tower of 
light particles can produce a distance divergence in a quantum corrected
metric. This field theory fact is consistent with our conjecture that 
an infinite distance singularity in the moduli space is 
correlated to an emergence of an infinite tower of light particles.  Note,
however, that we are not claiming that the infinite distance in moduli
space is a direct consequence of having an infinite tower of massless
states.  Indeed it is possible to show that these are independent conjectures
and one conjecture does not follow from the other.  For example,
the KK tower of light states (which can affect the distance 
in moduli space at one loop level) is not 
directly related to an infinite distance
in the moduli space which is observable at tree level.  Also sometimes
there is an infinite tower of light states but at finite distance. This
typically happens at conformal fixed points.  For example consider
M-theory compactified on a Calabi-Yau 3-fold near a point in moduli where a
${\bf P}^2$ inside the Calabi-Yau shrinks to a point.  
This leads to an infinite
tower of light M2 branes wrapping shrinking cycles in ${\bf P}^2$.  
Nevertheless
the distance in moduli space where ${\bf P}^2$ shrinks to zero size is finite.

\subsec{Observable Regions of Moduli Spaces}

For simplicity, let us consider the case when
there is one massless scalar field $\phi$ and choose
its parametrization so that the metric in the moduli space
is flat: $ds^2 = M_{Pl}^{-2} d\phi^2$, where $M_{Pl}$ is
the Planck mass. This means that the value of
$\phi$ is the geodesic length times $M_{Pl}$. 

Suppose $\phi$ is homogeneous in space. If it were not for gravity,
$\phi$ would evolve linearly in time $t$. If we turn on
the gravity and assume that the vacuum energy is dominated
by the kinetic term of $\phi$, we have the Freedman-Robertson-Walker
cosmology, where the Hubble friction slows down the scalar
velocity $\dot \phi = {d\phi\over dt}$:
$$ \ddot \phi  + dH \dot \phi = 0, ~~~
H^2 = {\dot \phi^2\over M_{Pl}^2},$$
where $d$ is the spatial dimensions of flat directions. 
With the initial condition,
$\dot \phi(t=0) = c$, we can integrate this to find
$$ \phi(t) = \phi(0) + {M_{Pl}\over d}  \log\left( 1
+ {c \over M_{Pl}} t\right).$$
In particular, $\phi(t)$ grows like $\log t$ for large $t$.

If this description were valid for all $t$, the scalar field
would continue to grow. However, the energy density of $\phi$
goes as $\dot \phi^2 \sim 1/t^2 \rightarrow 0$. Eventually,
other sources for $H$ become relevant. We should
then write,
$$ H^2 = {\dot \phi^2 \over M_{Pl}^2}  + H^2_{{\rm others}}
  \sim \left( {1\over d t}\right)^{-2}+ H^2_{{\rm others}}, $$
where $H_{{\rm others}}$ represents contributions from other sources. 
When $t \gg 1/H_{{\rm others}}$, the velocity of
the scalar field starts to decay exponentially,
$\dot \phi \sim \exp(-d\ H_{{\rm others}} t)$ and 
the evolution of $\phi$ is shut off. Thus we find that
$$ |\phi(t) - \phi(0)| \sim - {M_{Pl}\over d}\log H_{{\rm others}} ,
~~~t\rightarrow \infty. $$
If we identify $H_{{\rm others}}$ with the mass scale
$\epsilon$ for the low energy effective theory, this takes the same form as
the logarithmic cutoff $T < - M_{Pl} \log \epsilon$ for the geodesic
length in the moduli space. This suggests that the classical evolution
of the field $\phi$ cannot probe the region where the low energy
effective theory breaks down by the appearance of an infinite tower
of light particles. Similar ideas have been considered in
%\BanksES
\lref\Bankstwo{
  T.~Banks, M.~Dine and E.~Gorbatov,
  ``Is there a string theory landscape?,''
  JHEP {\bf 0408}, 058 (2004);
  {\tt hep-th/0309170}.
  %%CITATION = HEP-TH 0309170;%%
}
%\BanksVP
\lref\Banksone{
  T.~Banks,
  ``A critique of pure string theory: Heterodox opinions of diverse
  dimensions,''
  {\tt hep-th/0306074}.
  %%CITATION = HEP-TH 0306074;%%
}
\refs{\Banksone,\Bankstwo}.

 It is clear from the above discussion that the logarithmic
dependence of distance with the emergence of light mass scale
is very natural from the field theory
view point.  It would be nice to try to more clearly relate
the ideas from field theory with our conjectures about the appearance
of light modes at infinite corners of moduli space.

\newsec{Concluding Thoughts}

We have raised a number of conjectures about the geometry of scalar
moduli in a consistent quantum theory of gravity.  We find
determination of whether or not these and similar conjectures
are true or false as critical to a deeper understanding of what
constitutes the universality class of quantum gravitational theories.
Ultimately we would need to learn more:  Even if we
establish they are correct we would need to know {\it why}!  This
would presumably involve deep questions of gravity, such
as questions involved in the context of black hole.  It would be
very satisfying if we can relate such geometric questions on moduli
of scalar fields to deeper issues involving black holes and holography.
We hope raising such question will contribute to asking deeper questions
in the context of quantum gravity and how an effective field theory can
be possibly consistent with it.

\bigskip
\bigskip
\noindent {\bf Acknowledgments}
\bigskip

We would like to thank N. Arkani-Hamed for collaboration
at an early stage of this work.  In addition
we would like to thank D. Calegari, M. Douglas,
S. Kachru, Z. Lu, D. Morrison, L. Motl, A. Neitzke, H. Oh, M. Ro\v{c}ek, S. Shenker, 
A. Strominger, A. 
Van Proeyen, and S.-T. Yau for valuable discussions.
H.O. is grateful to the theory group of Harvard University
for their hospitality. H.O. also thanks
the organizers of the conference to celebrate the
60th birthday of Michael Green, where a preliminary 
version of this work was presented.  

 The research of H.O. is supported in part by
DOE grant DE-FG03-92-ER40701.
The research of C.V. is supported in part
by NSF grants PHY-0244821 and DMS-0244464.

\listrefs
\end
\bye